\documentclass[10pt,aps,pre,floatfix]{revtex4}
 \usepackage{xcolor}
\usepackage{graphicx,amssymb,amsfonts,amsmath,color,bm,hyperref}

\hypersetup{
    colorlinks,
    linkcolor={blue},
    citecolor={blue},
    urlcolor={black}
}

\begin{document}
\title{{\bf Work and heat distributions  for a Brownian particle 
subjected to an oscillatory drive}}
\author{Bappa Saha}
\author{Sutapa Mukherji}
\affiliation{Department of Physics, Indian Institute of Technology,
Kanpur-208 016}
\date{\today}
\begin{abstract}
Using the Onsager-Machlup functional integral 
approach, we obtain  the work distribution  function 
and the distribution of  the dissipated 
heat  of a  Brownian particle 
subjected to a confining harmonic 
potential and an oscillatory 
driving force.  
In the long time limit, the width of the 
work distribution function initially increases with the frequency of 
the driving force and  finally saturates  to a fixed value for  large values 
of the angular frequency.   Using the results from the work 
distribution part,  
we  next obtain the distribution of the dissipated 
heat for  the equilibrium initial condition. 
Using the method of steepest descent,
we obtain a Gaussian 
distribution for small fluctuations in the large time limit.  
The distribution function, for a fixed time has been obtained 
numerically.   It is shown that the heat distribution, in general, 
does not satisfy the transient fluctuation 
theorem.
\end{abstract}
\maketitle

\section{Introduction}
The fluctuation-dissipation theorem 
displaying the connection between the 
`friction coefficient' and  the fluctuations in thermodynamic 
variables has been derived for systems close to equilibrium 
\cite{reichl,kubo}. 
The role of fluctuations in nonequilibrium systems is described 
 through a set of  powerful, general results known as fluctuation 
theorems \cite{evans,gallavotti,kurchan,lebowitz,zon,zon_cohen1,zon_cohen2,
seifert,harris}. 
Fluctuation theorems have been proposed for various fluctuating 
thermodynamic quantities like heat, work,  entropy production etc and 
they  display the macroscopic irreversibility of the system subjected to 
nonequilibrium  conditions.  
According to the fluctuation theorem, the   probability $P(W_\tau=w\tau)$  
that a time-integrated 
quantity $W_\tau= \int_0^\tau {\rm d}t \ \dot{W}$ ( $W_\tau$ may 
represent  the work done by an external drive over time interval $\tau$) 
has a value 
$w\tau$ satisfies the following relation 
\begin{eqnarray}
\lim_{\tau \rightarrow \infty} \frac{1}{\tau}\ln 
\frac{P(W_\tau=w\tau)}{P(W_\tau=-w\tau)}
= w.
\end{eqnarray} 
The asymmetry displayed in this equation 
is due to the external field responsible for driving the 
system out of equilibrium. 
Fluctuation theorems are of two kinds:  the  
transient fluctuation theorem and 
the steady-state fluctuation theorem. In case of
 the transient fluctuation 
theorem, the  system evolves from an initial   equilibrium  state 
at $\tau=0$.  For the steady-state fluctuation theorem, the system is in 
its nonequilibrium stationary state through out the entire time
 interval $\tau$ \cite{evans}.

Onsager-Machlup used a functional integral approach  in their 
original study on  fluctuations in   linear-relaxation processes 
\cite{onsager1,onsager2}.
An outcome of a variational treatment on  the functional integral
 is Onsager's principle of minimum energy dissipation \cite{onsagerold}.
 Since the development of the  Onsager-Machlup fluctuation theory, 
there have been many 
efforts to  extend  their  functional integral 
approach to systems  far away from equilibrium
\cite{bertini1,bertini2,bertini3,taniguchi1,taniguchi2,cohen,maes}. 
Using the Onsager-Machlup approach for 
nonequilibrium steady states, Taniguchi and Cohen 
 obtained the work and heat  distribution functions
of a Brownian particle dragged by a moving 
harmonic potential \cite{taniguchi1,taniguchi2,cohen}.  
They showed the validity 
of the work fluctuation theorem in the long time limit for 
arbitrary initial conditions and discussed the extended heat fluctuation 
theorem  for the heat distribution function. 
Similar methods have also been 
used to investigate the asymptotics  of the work distribution 
functions for a number of nonequilibrium systems \cite{engel}.

In the present work, we extend the 
Onsager-Machlup functional integral approach 
to obtain the distributions 
of the total work and the dissipated heat  for a Brownian particle 
subjected to an external 
oscillatory driving force and a confining 
harmonic trap.  
We express the transition probability in a functional integral form 
in terms of a  Lagrangian.  The distribution of the 
 total work done on the particle
 can be expressed in terms of an appropriate 
functional integral involving the transition probability. 
Our approach involves an explicit evaluation of the functional 
integral to 
obtain the general form of the  distribution function. 
The distribution function  reflects clearly  the oscillatory features 
of the external drive. In the long time limit, 
the work distribution function becomes Gaussian with a 
variance that depends  non-trivially on the angular frequency of the 
oscillatory drive.  Analytical results  
appear  to be in agreement with those from numerical simulations.  
 Results from 
 the work distribution part   are  used further to obtain the Fourier 
transform of the distribution of the dissipated heat.   
Using the method of steepest descent, we 
have obtained a  Gaussian distribution 
 for the dissipated heat  over the central region (small fluctuations).  
The  general  behavior of the heat distribution, at a fixed time 
(not necessarily  large),
 is obtained by numerically
 evaluating the inverse Fourier transform. 
The numerically evaluated distribution for a finite  time  and 
the simulation results show that the distribution, 
in general,  does not satisfy the transient fluctuation theorem.
 
 Much before the application 
of the Onsager-Machlup theory 
to the nonequilibrium stationary state  problems involving a dragged Brownian particle 
\cite{taniguchi1,taniguchi2}, 
it has been shown 
in a general way that the  work distribution function for 
 a parabolic potential  with an arbitrary motion of the center 
 satisfies the transient and  stationary state fluctuation 
 theorem \cite{zon_cohen1,speck}. 
 The distribution function of heat for a 
  Brownian particle dragged by a moving parabolic  
  potential has been studied in 
\cite{zon_cohen2,imparato,taniguchi1,taniguchi2}.
 The  present  work   provides a detailed evaluation 
 of the work distribution and the  distribution of the dissipated heat 
  for a nonequilibrium oscillatory state using the 
 Onsager-Machlup  path integral approach. 
 Using this method,  the work distribution 
 for an oscillatory potential  has 
 been obtained  earlier in \cite{nsingh}.
 The  work distribution in \cite{nsingh} is essentially 
 for the mechanical 
 part of the work associated with the driving force.
 On the other hand, the work  
 distribution evaluated here  is for the  
 total  work which consists of the  change in the potential energy 
 as well as the dissipative part. We have further used this result 
   to find  the distribution 
 of the dissipated heat.  
 We obtain the Fourier transform of the heat distribution 
 whose Fourier inverse is  found out using the method 
 of steepest descent 
 under certain conditions and  also through  
 a direct numerical integration. We also independently obtain the 
 heat distribution function through numerical simulations.
 The numerical integration and the  simulation results 
  indicate,  in general, a  non-Gaussian 
 nature of the heat distribution function.

The paper is organized  in the following way. In section II, we 
introduce the Langevin equation describing the motion of the 
Brownian particle. Deriving the Fokker-Planck equation, 
we then obtain a functional integral 
description for the transition 
probability.   In section III, the explicit form of the 
 transition probability is determined 
using a variational approach. 
The functional  integral description of the 
transition probability involves an Onsager-Machlup type 
Lagrangian which is used later in section IV, 
to obtain  the entropy production 
rate  for this system. The 
energy-conservation principle further allows us 
to identify the  total work done, the 
rate of mechanical work, that contributes to the potential energy, and 
the dissipative part of the work \cite{zon_cohen1}.  
In section V, we obtain a general expression for  
the distribution function 
 of the total work.  Section VI provides a derivation 
of the distribution of the dissipated heat.  
 The details on the numerical simulations 
of our system are provided in section VII. 
We summarize our work  in Section VIII.
Some of the  details of the calculations related to the 
derivation of the transition probability,
and the distributions  are presented in  appendices.

\section{Brownian Particle in presence of an oscillatory driving force}

The over-damped Langevin equation describing the motion of 
the Brownian particle is given by 
\begin{eqnarray}
\alpha \frac{{\rm d}x_t}{{\rm d}t}=-\Xi\ 
\cos \omega t-k x_t+\zeta_t,\label{langevin}
\end{eqnarray}
where $\alpha$ is the friction coefficient and $\zeta_t$ is a 
Gaussian distributed noise arising due to the coupling of the system with a 
 thermal reservoir. The noise distribution is specified through the 
averages $\langle \zeta_t\rangle=0$ and 
$\langle\zeta_{t_1}\zeta_{t_2}\rangle=g\ \delta(t_1-t_2)$, 
with $g$ being the strength of the noise.
 The particle is subjected to a confining 
harmonic potential $\frac{k}{2} x_t^2$ and an oscillatory force 
of strength $\Xi$.   

Our aim here is to find out the transition probability 
$P(x_f,t_f|x_i,t_i)$ which describes the probability of finding the 
Brownian particle at position $x_f$ at time $t_f$ given that the 
particle is located at $x_i$ at time $t_i$. In order to 
determine this, we obtain the Fokker-Planck equation 
for  the probability distribution function $\rho(x,t)$ 
of finding the particle at $x$ at time $t$. 
The Fokker-Planck equation is 
\begin{eqnarray}
\frac{\partial \rho(x,t)}{\partial t}={\cal L}\rho(x,t),
\end{eqnarray}
where the Fokker-Planck operator ${\cal L}$ is 
\begin{eqnarray}
{\cal L}=\frac{\partial}{\partial x}\left[\left(\frac{kx}{\alpha}+
\frac{\Xi}{\alpha} \cos \omega t\right)+\frac{g}{2\alpha^2}
\frac{\partial}{\partial x}\right].
\end{eqnarray}
From now onwards, we use the following 
parameters $D=\frac{g}{2\alpha^2}$,
$\gamma=\frac{k}{\alpha}$ and 
$\eta=\frac{\Xi}{\alpha}$. 
The expression of $D$ can be further simplified by  using the 
fluctuation-dissipation theorem \cite{kubo} that leads to  
$g=2\alpha/\beta $ and  $D=1/(\alpha\beta)$.
The Fokker-Planck equation allows us to obtain 
the following  functional  integral description for 
the transition probability $P(x_f,t_f|x_i,t_i)$ \cite{wiegel,risken}
\begin{eqnarray}
P(x_f,t_f|x_i,t_i)=\int_{x_i}^{x_f} {\cal D}x_t \ 
\exp\left[\int_{t_i}^{t_f} {\rm d}t 
\ {\it L}(x_t,\dot{x}_t,t)\right],\label{pathintegral}
\end{eqnarray} 
where 
\begin{eqnarray}
{\it L}(x_t,\dot{x}_t,t)=-\frac{1}{4D}(\gamma x_t+\dot{x}_t+
\eta  \cos \omega t)^2. \label{lagrangian}
\end{eqnarray}
Here $\int_{x_i}^{x_f} {\cal D}x_t$ denotes a sum over all possible 
paths between the initial and final points,  $x_i$ and $x_f$, 
respectively.

In 
\begin{eqnarray}
\exp\left[-\frac{1}{4D} \int_{t_i}^{t_f}{\rm d}t\,(\gamma x_t+\dot{x}_t+
\eta \cos \omega t)^2\right]
\end{eqnarray} 
the integrand as well as the integral 
are  either zero or positive. 
The condition for zero of the integrand is satisfied 
by the average path which, from equation (\ref{langevin}),
 is the solution of 
\begin{eqnarray}
\frac{{\rm d}\langle x_t\rangle}{{\rm d}t}
=-\eta \cos \omega t-\gamma\langle x_t\rangle.
\end{eqnarray}
It is  equivalent to saying that at each time instant, one considers 
the average position of the Brownian particle. The path constructed 
this way is the average path. 
There is another special path which corresponds to 
a path in a given time interval $[t_i:t_f]$ with  maximum 
probability.  
This path is the  most probable path  that contributes maximally 
to the transition probability \cite{most-probable}.

\section{Transition probability}
The most probable path for the Brownian particle can be found out 
by extremizing the integral $\int_{t_i}^{t_f} dt\ 
{\it L}(x_t,\dot{x}_t,t)$ in equation (\ref{pathintegral}). The 
extremization leads to the Euler-Lagrange equation 
\begin{eqnarray}
\frac{{\rm d}}{{\rm d}t} \Big(\frac{\partial {\it L}}{\partial 
{\dot{x_t}}}\Big)-
\frac{\partial {\it L}}{\partial x_t}=0,
\end{eqnarray}
from which we find the 
following equation for the most probable  path
\begin{eqnarray}
\ddot{x}_t-\gamma^2 x_t-\eta 
(\omega\sin \omega t+\gamma\cos \omega t)=0,
\end{eqnarray}
where an overdot denotes a derivative with respect to time. 
The most probable path denoted 
as $\tilde{x}_t$ is found as 
\begin{eqnarray}
\tilde{x}_t=A \exp[\gamma t]+B\exp[-\gamma t] -
\frac{\eta \gamma}{\omega^2+\gamma^2} \cos\omega t-
\frac{\eta \omega}{\omega^2+\gamma^2} \sin\omega t,\label{optpath}
\end{eqnarray}
where $A$ and $B$ are the integration constants 
which depend on the initial conditions.
Using  the initial conditions, $\tilde{x}_t = x_i$ 
at $t=t_i=0$ and $\tilde{x}_t = x_f$  at  $t=t_f $, 
we find
\begin{eqnarray}
&&A = x_i+\frac{\gamma\eta}{\omega^2+\gamma^2}-
\frac{1}{e^{-\gamma t_f}-e^{\gamma t_f}} 
\left[ e^{\gamma t_f}\left(-\frac{\eta\gamma}{\omega^2+\gamma^2}-x_i\right)+
\frac{\eta\omega}{\omega^2+\gamma^2}\sin\omega t_f +
\frac{\eta\gamma}{\omega^2+\gamma^2}\cos\omega t_f +x_f\right], \\
&& B=x_i+\frac{\eta\gamma}{\omega^2+\gamma^2}-A.
\end{eqnarray}
The corresponding Lagrangian for the most probable path is given by 
  ${\it L}(\tilde{x}_t,\dot{\tilde{x}}_t,t)=
-\frac{1}{4D}(4\gamma^2 A^2 \exp[2\gamma t])$.

In order to obtain the explicit form of  $P(x_f,t_f|x_i,t_i)$,  we 
need to do the functional integration in (\ref{pathintegral}). 
The functional integration is done by considering paths with 
 infinitesimal deviations, $z_t$,  about the most probable path 
as $x_t=\tilde{x}_t+z_t$ and $\dot{{x}}_t=\dot{\tilde {x}}_t+\dot{z}_t$.
Expanding in small  $z_t$ and $\dot{z}_t$, we have 
\begin{eqnarray}
&&\int_{t_i}^{t_f} {\rm d}t\ {\it L}(\tilde{x}_t+z_t,\dot{\tilde{x}}_t+
\dot{z}_t,t) \nonumber\\
&&=\int_{t_i}^{t_f} {\rm d}t\ {\it  L}(\tilde{x}_t,\dot{\tilde{x}}_t,t)
+\int_{t_i}^{t_f} {\rm d}t\ \Big[
\frac{\partial{\it L}(\tilde{x}_t,\dot{\tilde{x}}_t,t)}
{\partial \tilde{x}_t}-\frac{{\rm d}}{{\rm d}t}
\frac{\partial {\it L}(\tilde{x}_t,\dot{\tilde{x}}_t,t)}
{\partial \dot{\tilde{x}}_t}\Big]z_t
-\frac{1}{4D}\int_{t_i}^{t_f} {\rm d}t\ 
\Big[\gamma^2 z_t^2+
\dot{z}_t^2+
2\gamma z_t \dot{z}_t\Big] \label{evaluation2}.
\end{eqnarray}
The term with a negative sign  in the 
second integral in expression 
(\ref{evaluation2}) is  obtained after  
doing  an integration by parts  of 
$\int_{t_i}^{t_f} {\rm d}t \ \frac{\partial {\it L}}
{\partial \dot{\tilde{x}}_t}(\tilde{x}_t,\dot{\tilde{x}}_t){\dot{z}_t}$ 
that appears at the first order in the  expansion. 
The transition probability can now be expressed as 
\begin{eqnarray}
P(x_f,t_f|x_i,t_i)=\exp\Big[\int_{t_i}^{t_f} {\rm d}t\ 
L(\tilde{x}_t,\dot{\tilde{x}}_t,t)\Big]
\int {\cal D}z \ \exp\Big[-\frac{1}{4D}\int_{t_i}^{t_f} {\rm d}t\ 
[\gamma^2 z_t^2+
\dot{z}_t^2+
{2\gamma}z_t \dot{z}_t]\Big].\label{evaluatepathint}
\end{eqnarray}
The functional integral in the above expression is to be done with the 
constraints $z_{t_i}=z_{t_f}=0$. Appendix \ref{App:AppendixA}
 provides the details 
of calculation of the  functional  integral in equation 
 (\ref{evaluatepathint}). 
The final result for the transition probability is 
\begin{eqnarray}
P(x_f,t_f|x_i,t_i)=\Big(\frac{2\pi D}{\gamma}\Big)^{-1/2}
\exp\Big[-\frac{\gamma A^2}{2D}(e^{2\gamma t_f}-
e^{2\gamma t_i})\Big] 
\Big(1-\exp[-2\gamma(t_f-t_i)]\Big)^{-1/2}.\label{transprob}
\end{eqnarray}

Clearly, at large time $t_f \rightarrow \infty$ , 
$A \approx (x_f+ \frac{\eta\omega}{\omega^2+\gamma^2} 
\sin \omega t_f + \frac{\eta\gamma}{\omega^2+\gamma^2}\cos \omega t_f) 
e^{-\gamma t_f}$. 
In this limit, the  transition probability has  the form,
\begin{equation}
P(x_f,t_f|x_i,t_i) = \left(\frac{2\pi D}{\gamma}\right)^{-1/2} 
\exp \left[-\frac{\gamma}{2D} \left(x_f+ 
\frac{\eta\omega}{\omega^2+\gamma^2} \sin \omega t_f + 
\frac{\eta\gamma}{\omega^2+\gamma^2}\cos \omega t_f\right)^2\right].
\end{equation}
The above result implies that in the absence 
of the oscillatory force ($\eta = 0$), one recovers, at large time, 
the  equilibrium
 probability distribution 
\begin{eqnarray}
\rho_{\rm eq}(x_f)&=&\int {\rm d}x_i\, P(x_f,t_f|x_i,t_i) f(x_i,t_i) 
\approx \left(\frac{2\pi D}{\gamma}\right)^{-1/2}
\exp \left[-\frac{\gamma x_f^2}{2D}\right]\nonumber\\
          &= & \left(\frac{2\pi }{k\beta}\right)^{-1/2}\exp 
\left[-\beta(k x_f^2/2)\right],
\end{eqnarray}
where we have assumed the  initial distribution, $f(x_i,t_i)$, 
to be  normalized, i.e., $\int f(x_i,t_i) \,{\rm d}x_i = 1$.

\section{Energy conservation and 
Onsager-Machlup Lagrangian}
 In section II, we have shown that the transition
 probability can be 
expressed  as a functional 
 integral  involving  an Onsager-Machlup type Lagrangian. 
The purpose of this section is to use Langevin equation  to obtain 
  a  formal expression for the  work done by the external force. 
 The Onsager-Machlup Lagrangian may be used to identify the rate 
of entropy production in this process.   We show that the 
 expressions  
for the work done by the external force and  the entropy 
production rate, as obtained here, consistently satisfy 
the energy conservation principle.
 
The Langevin equation expresses the force-balance condition, 
\begin{eqnarray}
(-\alpha \frac{{\rm d}x}{{\rm d}t}+\zeta)-\Xi \ \cos \omega t-k x=0,
\label{langevin-forcebalance}
\end{eqnarray}
where terms in the bracket represent the force on the 
particle due to the reservoir. For simplicity, in this subsection, 
we have removed 
the subscript of the variable $x$. 
Multiplying  with a small displacement of the particle, this  equation 
can be converted to the energy conservation equation
\begin{eqnarray}
{\rm d}{\cal Q}+{\rm d}U={\rm d}W,\label{internalenergy}
\end{eqnarray} 
where ${\rm d}U=\frac{\partial U(x,t)}{\partial x} {\rm d}x+
\frac{\partial U(x,t)}{\partial t} {\rm d}t$ is  an 
exact differential with 
$U(x,t)=\Xi \ x\ \cos\omega t+\frac{1}{2} k x^2$ as  the potential 
energy and 
${\rm d}W=\frac{\partial U(x,t)}{\partial t} {\rm d}t$ is the 
work done by the external 
force \cite{sekimoto}.  The term 
${\rm d}{\cal Q}=-(-\alpha \frac{{\rm d}x}{{\rm d}t}+\zeta) {\rm d}x$
is the heat released by the particle to the heat reservoir. 
This work $W$, referred to in this paper as the total work, thus 
consists of a dissipative part associated with the heat release 
and a mechanical part associated with the change in $U$.
Over the interval $\{t_i:t_f\}$, various quantities  
 can be written as 
\begin{eqnarray}
\Delta{\cal Q}=T\int_{t_i}^{t_f} \dot{S}\ {\rm d}t,\label{heat}\\
\Delta W=\int_{t_i}^{t_f} \dot{W} {\rm d}t=- \Xi \ 
\omega \int_{t_i}^{t_f} {\rm d}t\  x \sin\omega t, \label{work}\\
\Delta U=\int_{t_i}^{t_f} {\rm d}U=
(\frac{1}{2} k x^2 +\Xi \ x \cos\omega t)\mid_{t_i}^{t_f}, \label{energy}
\end{eqnarray} 
where  $\dot{W}=-\Xi \ x\  \omega\ \sin\omega t$.

The Onsager-Machlup Lagrangian   can be expressed as 
\begin{eqnarray}
&&  L=-\frac{1}{2 k_B}\biggl\{\left[\gamma^2 x^2+
2 \gamma\eta  x \cos \omega t\right]
+\left[ \dot{x}^2+
\eta^2  (\cos \omega t)^2\right]
-\left[-2 \gamma x \dot{x}-
2 \eta \dot{x} \cos\omega t\right]\biggr\}\frac{\alpha}{2T}\nonumber \\
&&=-\frac{1}{2k_B}\left\lbrace \Phi(\dot{x},t)+\Psi(x,t)-
\dot{S}(x,\dot{x},t)\right\rbrace ,
\end{eqnarray}
where 
\begin{eqnarray}
&&\Phi(\dot{x},t)= \frac{\alpha}{2T}\left(\dot{x}^2+
\eta^2 (\cos \omega t)^2\right),\\
&&\Psi({x},t)=\frac{\alpha}{2T} \left(\gamma^2 {x}^2+
2\gamma\eta x\cos\omega t\right)
\end{eqnarray}
are the dissipation functions and 
\begin{eqnarray}
\dot{S}(x,\dot{x},t)=-2 \left(\frac{k}{\alpha}\right) x 
\dot{x}\frac{\alpha}{2T}-
2 \left(\frac{\Xi}{\alpha}\right)
\left(\frac{\alpha}{2T}\right)\dot{x} \cos\omega t 
\label{entropyprod}
\end{eqnarray}
is the rate of entropy production. 
Using (\ref{entropyprod}), (\ref{heat}) and (\ref{work}), one finds 
the change in the energy as expressed in (\ref{energy}).

\section{Work Distribution}\label{sec:workd}
We now determine the work distribution $P(W,t)=
\langle\langle\delta\big(W-{\cal W}(\{x_t\})\big)\rangle\rangle$, where 
${\cal W}(\{x_t\})=\int_{t_i}^{t_f} {\rm d}t\,
\frac{\partial U(x_t,t)}{\partial t}$ denotes 
the work done by the time-dependent oscillatory force 
along the path $x_t$ in 
time interval $[t_i:t_f]$
and $P(W,t)$ implies the probability  that this work has a value $W$. 
$\langle\langle--\rangle\rangle$ denotes a
functional average over all possible paths in the given 
time interval and integrals
over initial and final positions.

We express $P(W)$ in terms of the Fourier transform of 
${\cal W}(\{x_t\})$ as 
\begin{eqnarray}
P(W)=\frac{1}{2\pi} \int_{-\infty}^{\infty} {\rm d}\lambda\  
e^{i\lambda W}
\langle\langle e^{-i\lambda {\cal W}(\{x_t\})}\rangle\rangle.
\label{finaldist1}
\end{eqnarray}
 The functional integral can be expressed in terms of the 
 transition probability 
 shown in (\ref{pathintegral}).  Since the  
Lagrangian function  $L$ 
  in the transition probability has a prefactor  $\beta$, 
  it is convenient if we introduce a $\beta$ in equation 
  (\ref{finaldist1}) and re-express it as,
\begin{eqnarray}
P(W)=\frac{\beta}{2\pi} \int_{-\infty}^{\infty} {\rm d}\lambda\  
e^{i\lambda \beta W}
\langle\langle e^{-i\lambda \beta{\cal W}(\{x_t\})}\rangle\rangle. 
\label{finaldist2}
\end{eqnarray}
Thus 
\begin{eqnarray}
&&\langle\langle e^{-i\lambda\beta {\cal W}(\{x_t\})}\rangle\rangle
=\int {\rm d}x_f \int {\rm d}x_i \ f(x_i,t_i) \int_{x_i}^{x_f} {\cal D} x_t\  
e^{-i \lambda \beta {\cal W}(\{x_t\})} e^{\int_{t_i}^{t_f} {\rm d}t 
{\it L}(x_t,\dot{x}_t,t)}\nonumber\\
&&=\int {\rm d}x_f \ \int {\rm d}x_i\ f(x_i,t_i)\ {\cal F}(x_f,x_i,\lambda),
\label{average}
\end{eqnarray}
where 
\begin{eqnarray}
{\cal F}(x_f,x_i,\lambda)=\int_{x_i}^{x_f} {\cal D}x_t\ 
 e^{\int_{t_i}^{t_f}{\rm d}t\ \Big[{\it L}(x_t,\dot{x}_t,t)-
i\beta\lambda \dot{\cal W}(x_t)\Big]}.\label{pathintwork0}
\end{eqnarray}
Here the functional average is done over all possible paths 
extending from $x_i$ to $x_f$ over the time interval $[t_i:t_f]$.
The final result is  obtained upon  averaging 
over all initial points with the distribution $f(x_i,t_i)$ and 
 integrating over the  final point. 
As discussed before, in our case 
 $\dot{\cal W}(x_t)=\frac{\partial U}{\partial t}=-\Xi \  
x_t\ \omega\ \sin(\omega t)$. 
The functional integral in (\ref{pathintwork0})
is evaluated by maximizing  the integral $\int_{t_i}^{t_f} 
{\rm d}t\ \Big[{\it L}(x_t,\dot{x}_t,t)-
i\lambda \beta\dot{\cal W}(x_t)\Big]$. This 
 leads to the  Euler-Lagrange equation 
\begin{eqnarray}
\frac{{\rm d}}{{\rm d}t} \Big(\frac{\partial {\it L}(x_t,\dot{x}_t,t)}
{\partial {\dot{x}}_t}\Big)-
\frac{\partial {\it L}(x_t,\dot{x}_t,t)}{\partial x_t}+i\beta\lambda 
\frac{\partial \dot{\cal W}(x_t)}{\partial x_t}=0.
\end{eqnarray} 
The   Euler-Lagrange equation written in terms 
of $x_t$, 
\begin{eqnarray}
\ddot{x}_t-\gamma^2 x_t
-(\eta \omega-i(2D\lambda \ \beta\ \Xi\ \omega)) \sin\omega t-
\eta\gamma \cos \omega t=0
\end{eqnarray}
has a solution of the form 
\begin{eqnarray}
x_t^*=A_W e^{\gamma t}+B_W e^{-\gamma t} -
\frac{(\eta\omega-i\bar{\lambda})}{\omega^2+\gamma^2}\sin\omega t
-\frac{\eta\gamma}{\omega^2+\gamma^2}\cos\omega t,
\label{mostprobableW}
\end{eqnarray}
where $\bar{\lambda}=2D\lambda\beta \Xi\omega=
2\lambda\eta\omega$.
The two integration constants,  $A_W$ and $B_W$,  determined using  
the initial and final conditions, $x_t^*=x_{f},\ {\rm at}\ t=t_f$
and $x_t^*=x_i$ at $t=t_i=0$, are 
\begin{eqnarray}
 && A_W=\left[x_{f}-\left(x_i+\frac{\eta \gamma}{(\omega^2+\gamma^2)}\right)
e^{-\gamma t_f}
+\frac{\eta( \omega \sin\omega t_f+\gamma\cos\omega t_f)}
{\omega^2+\gamma^2}-\frac{i\bar{\lambda}\sin\omega t_f}
{\omega^2+\gamma^2}\right] 
(\exp[\gamma t_f]-\exp[-\gamma t_f])^{-1}\label{aw} \\ 
&& \  \ {\rm and} \ \ 
B_W=x_i+\frac{\eta \gamma}{\gamma^2+\omega^2}-A_W.\label{bw}
\end{eqnarray}
The quantity ${\it L}(x_t^*,\dot{x_t}^*)-i 
\lambda \beta \dot{\cal W}(x_t^*)$ now has a form 
\begin{eqnarray}
{\it L}(x_t^*,\dot{x_t}^*)-i\lambda\beta\dot{\cal W}(x_t^*)
=-\frac{1}{4D}\left[2\gamma A_W e^{\gamma t}+
\frac{i\bar{\lambda}}{(\omega^2+\gamma^2)} (\gamma\sin\omega t
+\omega\cos\omega t)\right]^2+\frac{i\bar{\lambda}}{2D} x_t^* \sin\omega t.
\end{eqnarray}
As before, the evaluation of ${\cal F}(x_f,x_i,\lambda)$ is done 
about the path $(x_t^*,\dot{x}_t^*)$ which maximizes  
$\int_{t_i}^{t_f} 
{\rm d}t\ \Big[{\it L}(x_t,\dot{x}_t,t)-
i\beta\lambda \dot{\cal W}(x_t)\Big]$. Thus 
\begin{eqnarray}
{\cal F}(x_i,x_f,\lambda)=e^{\int_{t_i}^{t_f} {\rm d}t\ 
[ {\it L}(x_t^*,\dot{x_t}^*)
-i\lambda \beta\dot{\cal W}(x_t^*)]}
\int {\cal D}z_t \ e^{-\frac{1}{4D}\int_{t_i}^{t_f} {\rm d}t\ (\gamma z_t+
\dot{z_t})^2},\label{pathintwork}
\end{eqnarray}
where $z_t$ describes a small deviation about the maximal path as 
$x_t=x_t^*+z_t$ and $\dot{x_t}=\dot{x_t}^*+\dot{z_t}$ with 
the condition that $z_t$ vanishes at the initial and final 
time points. The derivation 
of equation (\ref{pathintwork}) is similar to 
 that done before for 
the transition probability. 
The functional integral  $\int {\cal D}z_t \ 
e^{-\frac{1}{4D}\int_{t_i}^{t_f}{\rm d}t \ (\gamma z_t+\dot{z}_t)^2}$ has
 been evaluated earlier
(see Appendix A).
Using this result,
we find 
\begin{eqnarray}
{\cal F}(x_f,x_i,\lambda)=\Big(\frac{2\pi D}{\gamma}\Big)^{-1/2} 
(1-e^{-2\gamma t_f})^{-1/2} e^{-\frac{1}{4D}(x_f^2 c_1+x_f c_2+c_3)}.
\label{beforexfint}
\end{eqnarray}
The expressions for $c_1, c_2, c_3$ are given in Appendix \ref{App:AppendixB}.

\begin{figure}[!ht]
  \begin{center}
  (a) \includegraphics[width=.45\textwidth,clip,
angle=0]{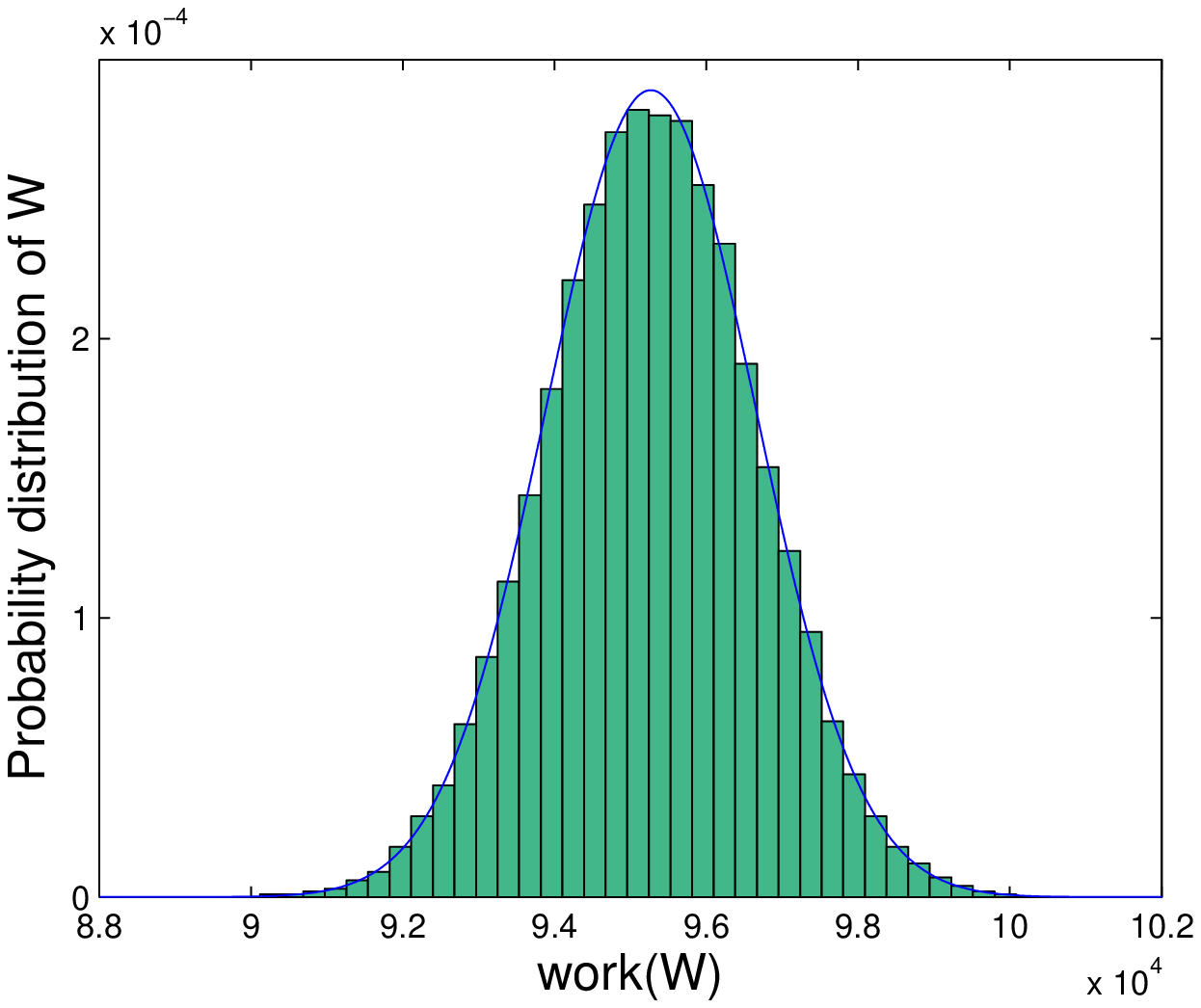}
(b)   \includegraphics[width=.45\textwidth,clip,
angle=0]{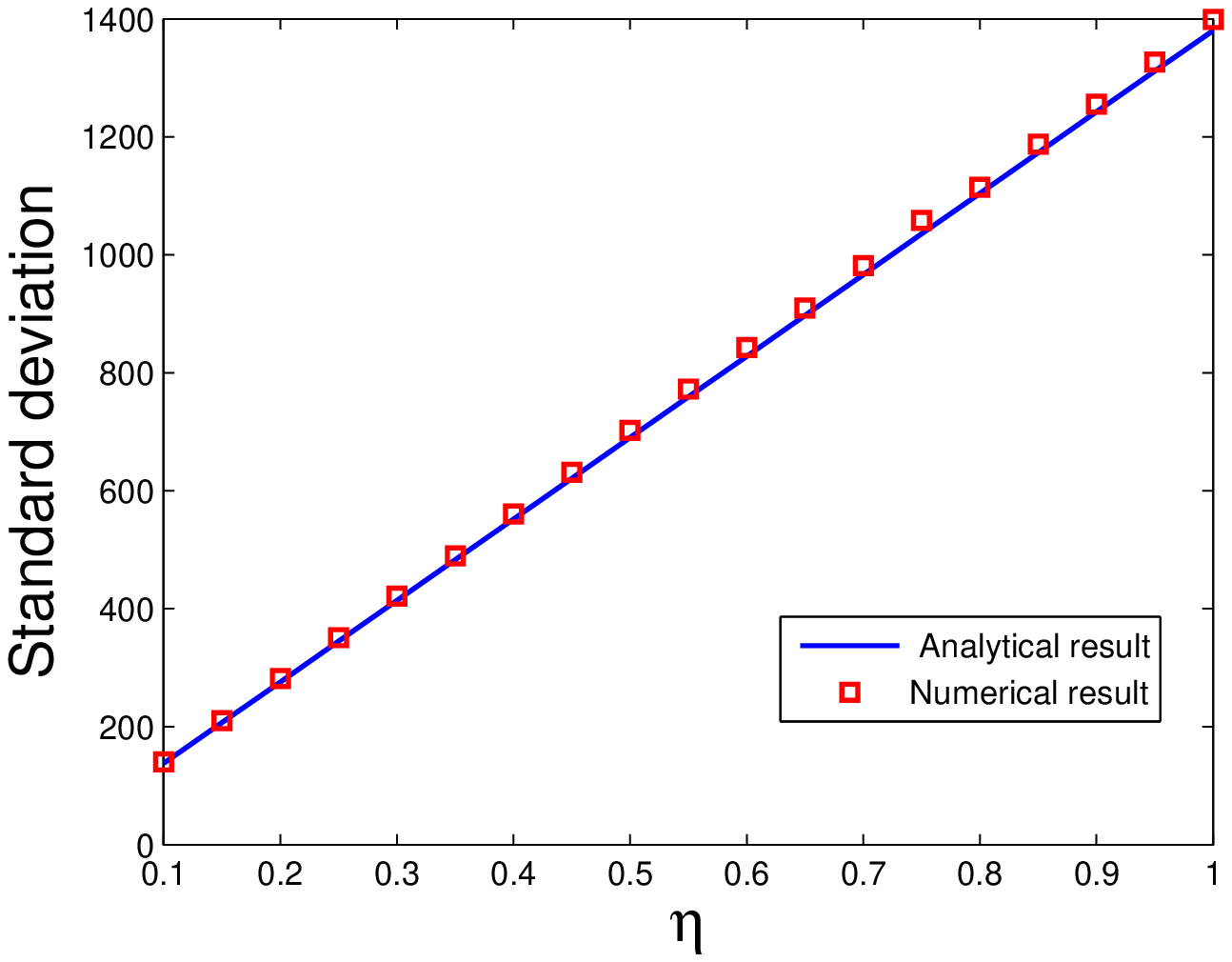}
\caption{(a) Work distribution at large time $t_f=1000$.
 The solid blue curve is the plot of the analytical 
result given by equation(\ref{largetimedist}). The parameter values 
are $D=0.05, \omega=0.1\pi, \gamma=0.07, \beta = 0.1, 
\alpha = 200 , \eta = 1$ . The histogram is obtained 
from numerical simulations, where, the final displacement and 
the work values are  computed for $40000$ number of 
realizations with time-step, $\Delta t= 0.05$.  
(b) Variation of the width of the work 
distribution with the strength of 
the oscillatory force $\eta$.
The work values are calculated for final time  $t_f=1000$. The solid 
blue line  represents the  standard deviation 
given by equation (\ref{stdev}) and the red squares correspond to 
the numerical results. The other parameter values are 
$D=0.05, \omega=0.1\pi, \gamma=0.07, \beta = 0.1, \alpha = 200$.  
The small time interval is 
$\Delta t= 0.25$.  For each $\eta$, 
the work values are computed for $30000$ realizations.}
\label{fig:workdist_lt}
  \end{center}
\end{figure}

Combining equations (\ref{finaldist2}) and (\ref{average}), we may write 
\begin{eqnarray}
P(W)=\frac{\beta}{2\pi} \int {\rm d}x_i \ f(x_i,t_i) 
\int_{-\infty}^{\infty} {\rm d}\lambda\ e^{i\lambda\beta W}
\int_{-\infty}^{\infty} {\rm d}x_f\  {\cal F}(x_f,x_i,\lambda).
\end{eqnarray}
Using (\ref{beforexfint}), and doing a Gaussian integration over $x_f$,
we obtain 
\begin{eqnarray}
P(W)=\left(\frac{2\gamma}{c_1}\right)^{1/2} 
(1-e^{-2\gamma t_f})^{-1/2} 
\frac{1}{2\eta\omega} \left(\frac{\beta}{2\pi}\right)
\int {\rm d}x_i\  f(x_i,t_i) \int_{-\infty}^{\infty} {\rm d}\bar{\lambda} \ 
e^{-(m_2{\bar\lambda}^2+m_1{\bar\lambda})} e^{i\bar\lambda\bar W},\label{m1m2}
\end{eqnarray}
where $\bar W=\frac{W}{2D\ \Xi\ \omega}$. 
The expressions for $m_1$ and $m_2$ are also given 
in Appendix \ref{App:AppendixB}.

After completing the integration over $\bar{\lambda}$, we have 
\begin{eqnarray}
P(W)=\left(\frac{\gamma\beta^2}{2\pi c_1m_2}\right)^{1/2} \ 
\frac{(1-e^{-2\gamma t_f})^{-1/2}}{2\eta \omega}
\int {\rm d}x_i \ f(x_i,t_i)\ 
\exp\left[-\frac{(\bar{W}-W_m)^2}{4 m_2}\right], \label{finaldist}
\end{eqnarray}
where $W_m=m_1/i$.

\begin{figure}[!ht]
  \begin{center}
  (a) \includegraphics[width=.45\textwidth,clip,
angle=0]{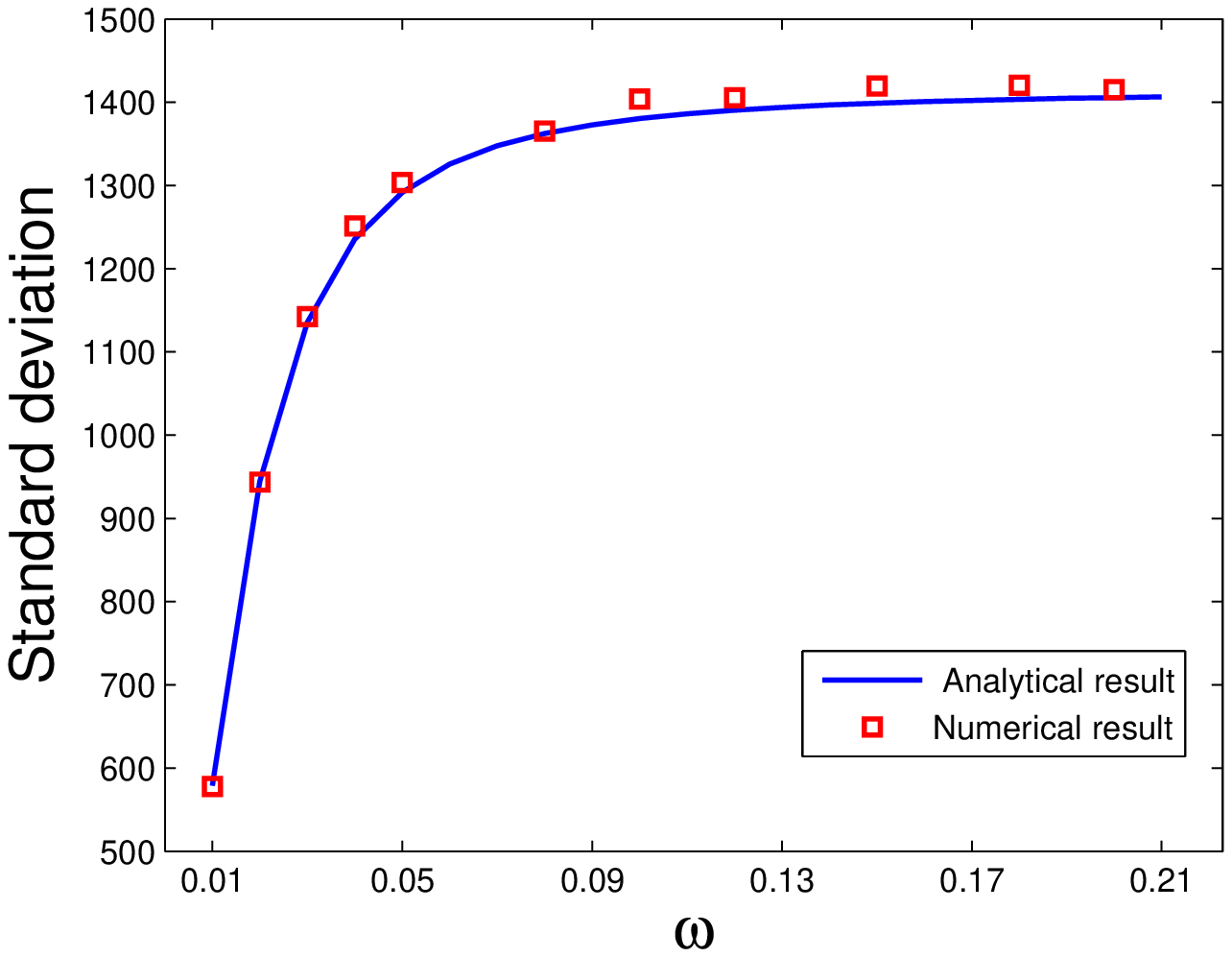}
(b)   \includegraphics[width=.45\textwidth,clip,
angle=0]{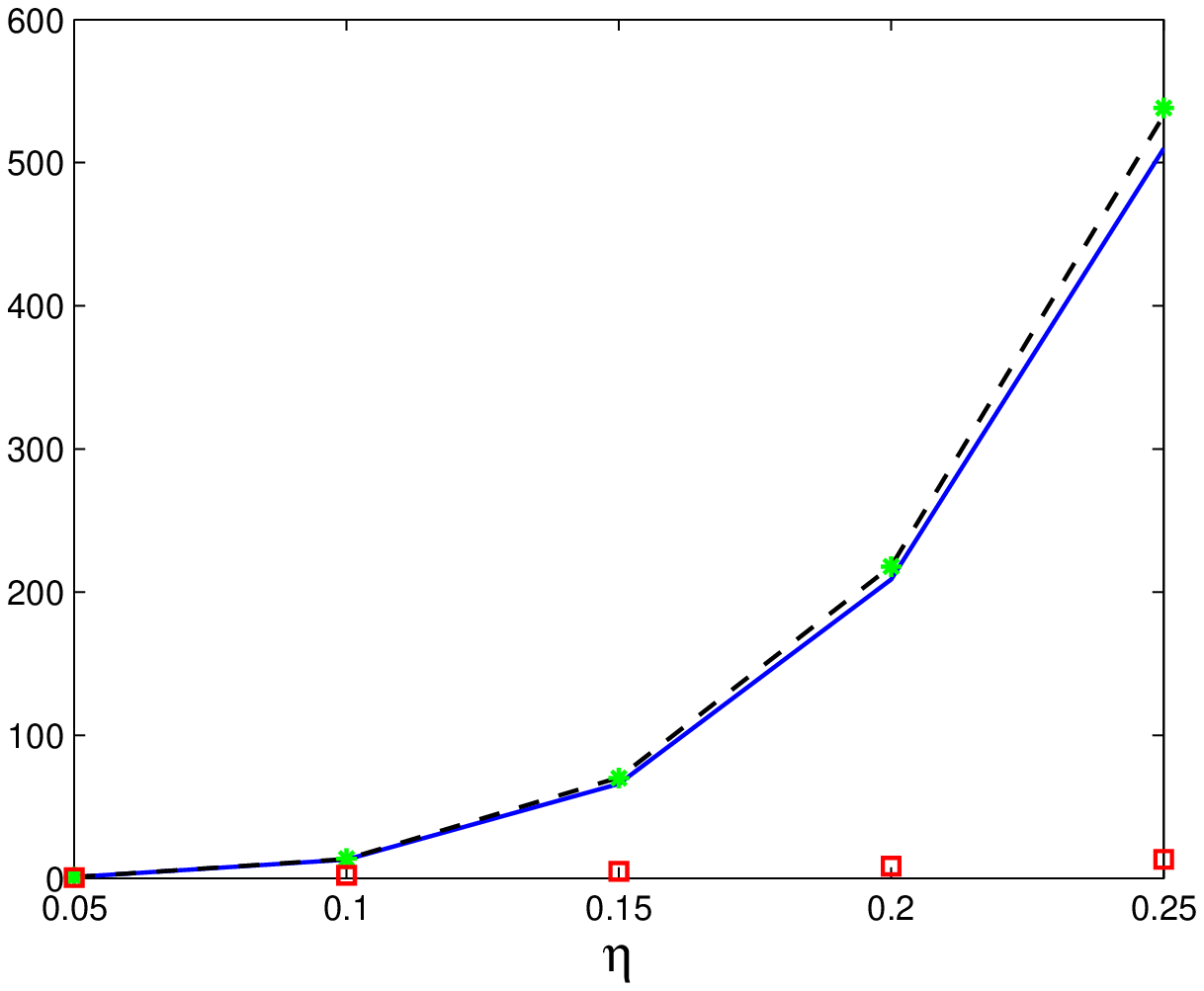}
\caption{(a) Variation of the width of the work distribution with the 
angular frequency $\omega$ (in the unit of $\pi$) of the
oscillatory force at large time $t_f=1000$. The solid blue  curve 
corresponds to the analytical value of the standard deviation given by  
equation (\ref{stdev}) and the red squares
correspond to numerical results. The other parameter values are $D=0.05,\  
\gamma=0.07,\ 
\beta=0.1, \  \alpha=200,\  \eta=1 $.  The small time interval is 
$\Delta t = 0.08 $. For each $\omega$,
the work-values have been computed for $30000$ realizations.  
(b) Plot of the variance and the central fourth moment 
of the work distribution 
with the strength, $\eta$,  of the oscillatory force at time $t_f=1000$. 
The solid blue line is obtained from equation(\ref{fourthcm}) and the 
black dashed line is the numerical value of 
$3(\langle W^2\rangle-\langle W\rangle^2)^2$. The green stars and 
the red squares are the simulation results of 
$\langle(W-\langle W\rangle)^4\rangle$ and 
$(\langle W^2\rangle-\langle W\rangle^2)$, respectively. 
The other  parameter values are $D=0.01, \omega=0.05\pi, 
\gamma=0.07, \beta=20, \Delta t=0.2$. 
All data are obtained after  averaging 
over $50000$ realizations. }
\label{fig:sd_freq}
  \end{center}
\end{figure}

In the large time limit ($t_f\rightarrow \infty$), the distribution 
function (\ref{finaldist}) becomes independent of the initial 
distribution $f(x_i,t_i)$($\int dx_i f(x_i,t_i)=1$). 
In this  limit, the 
work distribution has  the form 
\begin{eqnarray}
{\lim_{t_f\to \infty}} 
P(W,\ t_f)\sim\frac{\beta}{2\eta\omega}\frac{1}{(2\pi \sigma^2)^{1/2}} 
\exp\left[-\frac{({\bar W}-\eta \omega\sigma^2)^2}{2\sigma^2}\right],
\label{largetimedist}
\end{eqnarray}
where $\sigma^2=\frac{t_f}{4D(\omega^2+\gamma^2)}$.   
A similar Gaussian distribution  in the long time limit was found  
for a dragged Brownian particle  in \cite{mazonka,speck,taniguchi1}.
Thus, in the long time $(t_f\rightarrow \infty)$ limit,
the work distribution function 
satisfies the work fluctuation theorem,
\begin{eqnarray}
\lim_{t_f\to \infty} \frac{P(W,t_f)}{P(-W,t_f)}=\exp[\beta W]
\end{eqnarray}  
 for  any  initial 
condition.

In the present problem of a nonequilibrium  oscillatory state, 
 the width of the distribution 
\begin{eqnarray}
(\langle W^2\rangle-\langle W\rangle^2)^{1/2}=
\frac{2\eta \omega \sigma}{\beta} \label{stdev}
\end{eqnarray}
 increases linearly with the strength of the  oscillatory 
force.  Further,  for a given 
 strength of the oscillatory force,  the width increases 
initially with the 
 angular frequency $\omega$ 
 and  finally saturates to a value that depends on the strength 
 of the oscillatory drive. 
 The fourth central moment of the work distribution in equation 
 (\ref{largetimedist}) is 
 \begin{eqnarray}
 \langle(W-\langle W\rangle)^4\rangle  = 
\frac{48 \eta^4 \omega^4 \sigma^4}{\beta^4}. \label{fourthcm}
 \end{eqnarray}

In the following section, we derive  
the distribution of heat using the results obtained so far.

\section{Distribution of dissipated heat}\label{sec:heatd}
The dissipated heat can be defined as
\begin{eqnarray}
{\cal Q}={\cal W}(x_t)-\Delta U, \label{heatdef}
\end{eqnarray}
where  $\Delta U=U(x_f,t_f)-U(x_i,t_i)$ and
 $U=\frac{1}{2}kx^2+\Xi\ x\cos(\omega t)$.
The distribution of heat can be expressed as 
\begin{eqnarray}
P(Q)=\frac{\beta}{2\pi}\int_{-\infty}^{\infty} 
\mathrm{d}\lambda \,e^{i\lambda\beta Q}\langle\langle 
e^{-i\lambda\beta {\cal Q}}\rangle\rangle .\label{heatdist}
\end{eqnarray}
We denote $\langle\langle e^{-i\lambda\beta {\cal Q}}\rangle\rangle$
 by $P(i\lambda {\cal Q},t_f )$ and it can be determined by 
evaluating the functional averages over all possible paths 
as well as the integrals over initial and final positions.
Hence,
\begin{eqnarray}
P(i\lambda {\cal Q},t_f )= \int \mathrm{d}x_f\int 
\mathrm{d}x_i\, f(x_i,t_i) \ e^{i\lambda\beta(U(x_f,t_f)-U(x_i,t_i))}
{\cal F}(x_f,x_i,\lambda) ,
\end{eqnarray}
where, we directly substitute the expression of
 ${\cal F}(x_f,x_i,\lambda)$ from equation(\ref{beforexfint}).

Assuming that  the system evolves from initial equilibrium distribution, 
we express the normalized initial distribution of
 position $f(x_i,t_i=0)$ as 
\begin{eqnarray}
f(x_i)=\sqrt{\frac{\gamma}{2\pi D}}\ 
e^{-\frac{\eta^2}{2D\gamma}}\ e^{-(\frac{1}{2}\gamma x_i^2 +\eta x_i)/D} .
\end{eqnarray}
The shifted Gaussian nature of the distribution follows from the confining 
harmonic potential (defined after equation (\ref{internalenergy})) with its 
 minimum varying in an oscillatory fashion.
After performing the integrations over $x_f$ and $x_i$, 
we write the final form of $P(i\lambda {\cal Q},t_f )$ 
as a function of $\lambda$,
\begin{eqnarray}
P(i\lambda {\cal Q},t_f )=\frac{e^{-\eta^2/(2D\gamma)}}
{\left(1+\lambda^2 -e^{-2\gamma t_f}\lambda^2\right)^{1/2}}\ 
\exp(q_n/q_d) , \label{fouriert}
\end{eqnarray}
where, $q_d$ and $q_n$ are given by,
\begin{eqnarray}
q_d=4 D\gamma e^{2\gamma t_f}\ (1+\lambda^2- e^{-2\gamma t_f}
\lambda^2)\ (\omega^2+\gamma^2)^2
\end{eqnarray}
and 
\begin{eqnarray}
q_n= l_4\lambda^4+ i l_3\lambda^3 + l_2\lambda^2+
 i l_1\lambda+l_0 ,\label{exprqn}
\end{eqnarray}
where the expressions for $l_4,l_3,l_2,l_1$ and $l_0$ 
are given in Appendix \ref{App:AppendixB}.
Substituting equation(\ref{fouriert}) into (\ref{heatdist}), we have
\begin{eqnarray}
P(Q)=\frac{\beta}{2\pi}\int_{-\infty}^{\infty} \mathrm{d}\lambda \,
e^{i\lambda\beta Q}\frac{e^{-\eta^2/(2D\gamma)}}{\left(1+\lambda^2 
-e^{-2\gamma t_f}\lambda^2\right)^{1/2}}\ \exp(q_n/q_d) .\label{heatdist_ft}
\end{eqnarray}

In order to obtain an approximate
 form of the distribution in the 
 $t_f\rightarrow\infty$ limit, 
 equation (\ref{heatdist_ft}) can be written as,
\begin{eqnarray}
P(Q)=\frac{\beta}{2\pi}e^{-\frac{\eta^2}{2D\gamma}}
\int_{-\infty}^{\infty}\frac{\exp(t_f\ S(\lambda))}{(1+\lambda^2)^{1/2}}\, 
\mathrm{d}\lambda. \label{steepest_des}
\end{eqnarray}
where, 
\begin{eqnarray}
S(\lambda)=i\lambda\beta Q_f-\frac{\omega^2\eta^2
(\lambda^4+i\lambda^3+\lambda^2+i\lambda)}
{2D(\omega^2+\gamma^2)(1+\lambda^2)},
\end{eqnarray}
and $Q_f=Q/t_f$.
Since, in the  $t_f\rightarrow\infty$ 
limit, the exponent in the 
integrand in equation (\ref{steepest_des}) is linear in $t_f$, 
the integral can be evaluated using the method of steepest 
descents \cite{arfken}.  In this method, the integral is approximated 
by  the largest contribution that  comes from the saddle point.
 The saddle 
point is determined by extremizing  $S(\lambda)$. 
After finding out the saddle points, 
we deform the contour for $\lambda$ in the complex plane in such a way  
that it passes through the saddle point and the saddle point 
corresponds to the maximum of $S(\lambda)$ along the path. 

\begin{figure}[!ht]
  \begin{center}
  (a) \includegraphics[width=.45\textwidth,clip,
angle=0]{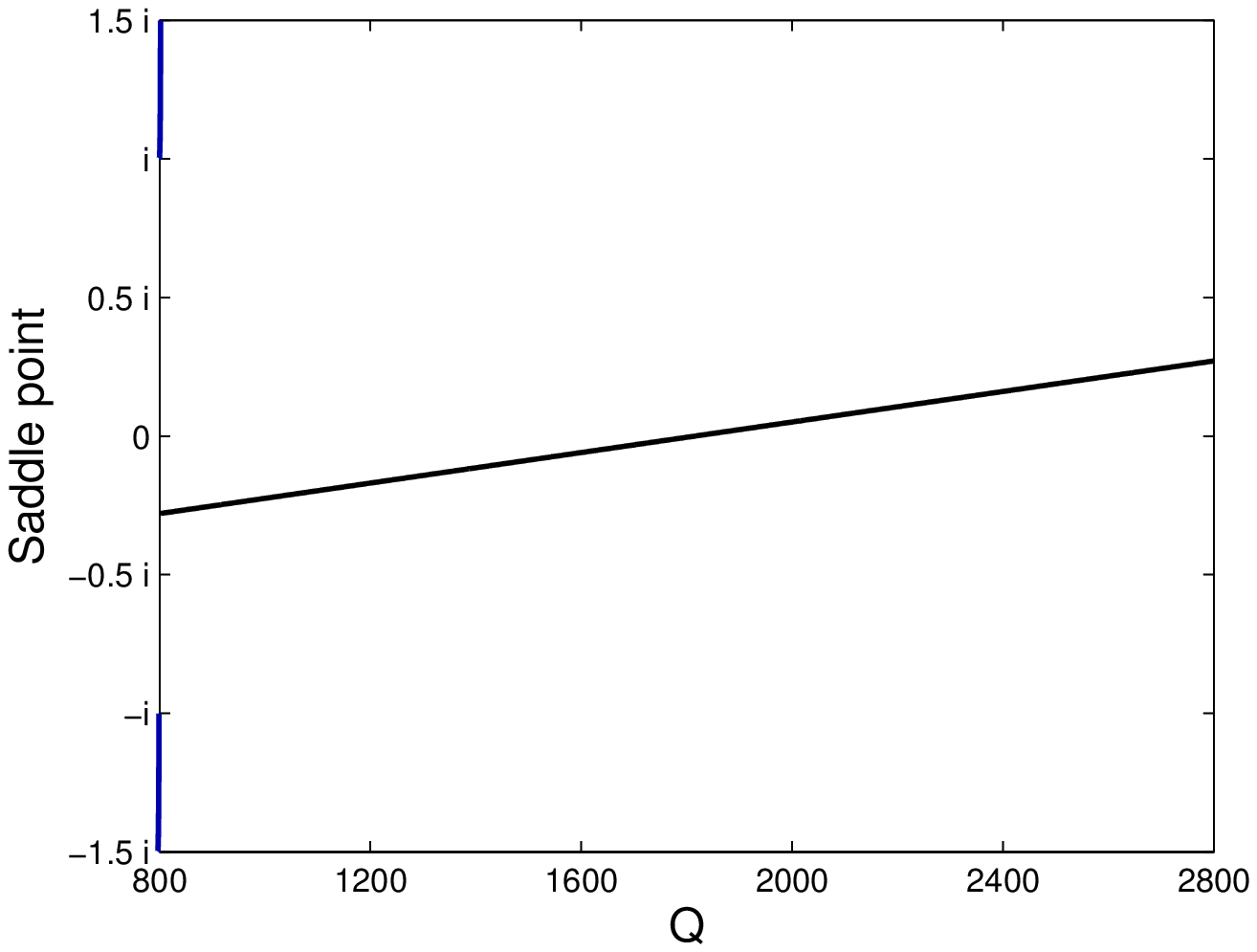}
(b)   \includegraphics[width=.45\textwidth,clip,
angle=0]{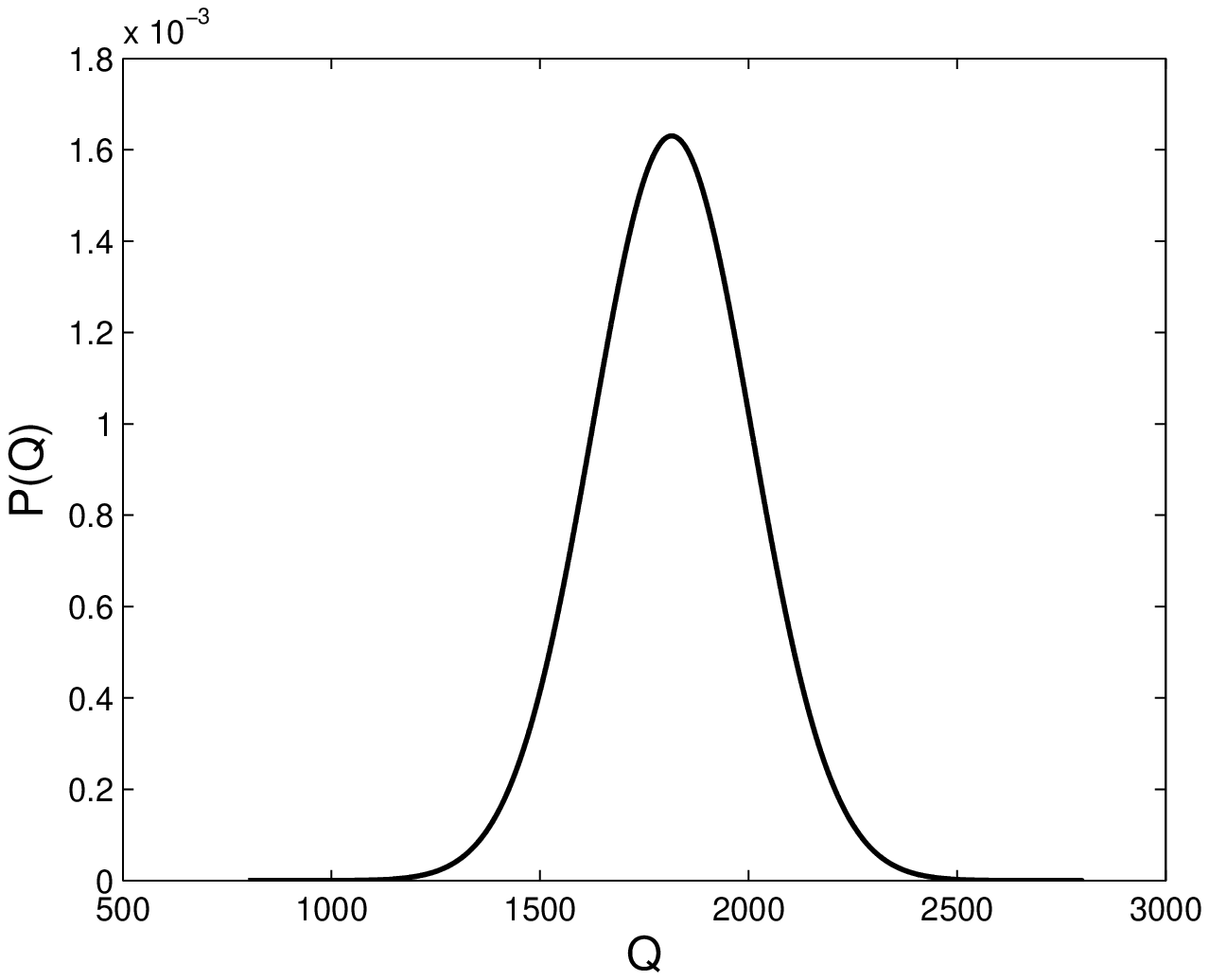}
\caption{(a) Plot of the saddle point with $Q$. 
We choose the range of $Q$ such that the saddle 
point is  in the range of  $[-i:i]$. The bold blue 
lines above $i$ and below $-i$ represent the branch cuts. 
The parameter values for this plot are
$D=0.05,\ \eta=0.05,\ \omega=0.1\pi,\ \gamma=0.1,\ \beta=0.1,
\ \alpha=200$ and $t_f=8000$.  
(b) The asymptotic distribution of the dissipated heat 
at long time $t_f=8000$ (see relation(\ref{finaltdist})). 
Other parameter values are same as those mentioned in (a). }
\label{fig:heatpdf_largetime}
  \end{center}
\end{figure}

Since the derivative of $S(\lambda)$ with respect to $\lambda$ is  
\begin{eqnarray}
S'(\lambda)=i\beta Q_f-\frac{\omega^2\eta^2}{2D(\omega^2+\gamma^2)}(2\lambda+i),
 \end{eqnarray} 
 there is only one saddle point $\lambda_0$ 
\begin{eqnarray}
\lambda_0= \frac{i(2\beta D Q_f(\omega^2+\gamma^2)-\omega^2\eta^2)}
{2\omega^2\eta^2} \label{saddlepoint}
\end{eqnarray}
located on the imaginary axis.  We further note that there are two 
branch points 
located at $\pm i$, and we choose the 
imaginary axis extending from $i$ to $+\infty$ and $-i$ to $-\infty$ 
to be the branch cut and the contour should not cross these cut lines. 
After  doing a Taylor expansion and  a Gaussian 
integral, we obtain the leading term of the heat distribution as 
\begin{eqnarray}
P(Q)\sim \frac{\beta}{2\pi}\ e^{-\frac{\eta^2}{2D\gamma}}
\left(\frac{2\pi}{t_f|S^{\prime\prime}(\lambda_0)|}\right)^{1/2}
\frac{\exp(t_f S(\lambda_0))}{(1+\lambda_0^2)^{1/2}}\ e^{i\theta} .\label{dist1}
\end{eqnarray} 
where, double prime denotes two  derivatives  with respect to 
 $\lambda$. $\theta$ gives the direction of the 
steepest descent and, in this case, 
 it is given by 
\begin{eqnarray}
\theta=\frac{\pi}{2}-\frac{1}{2}{\rm arg}[S^{\prime\prime}(\lambda_0)]=0 
\end{eqnarray}
which indicates a contour parallel to the real axis. 
In order to avoid crossing the branch cuts, we restrict $Q_f$ such that 
the saddle point is always  located between $i$ and $-i$.
The variation in the saddle point with $Q$ for a given set of 
 parameter values is shown in figure(\ref{fig:heatpdf_largetime}-a). 
Upon substituting $\lambda_0$, $S(\lambda_0)$ 
and $|S^{\prime\prime}(\lambda_0)|$ into equation (\ref{dist1}), 
the final form of the distribution, in the large  time limit, becomes
\begin{eqnarray}
P(Q)= \beta\ \left(\frac{D(\omega^2+\gamma^2)}
{2\pi t_f\omega^2\eta^2}\right)^{1/2}\frac{e^{-\frac{\eta^2}{2D\gamma}}}
{\left[1-\frac{(2\beta D Q_f(\omega^2+\gamma^2)-\omega^2\eta^2)^2}
{4\omega^4\eta^4}\right]^{1/2}}\ 
\exp\left[\frac{-t_f(2\beta D 
Q_f(\omega^2+\gamma^2)-\omega^2\eta^2)^2}
{8D\omega^2\eta^2(\omega^2+\gamma^2)}\right] \label{finaltdist}.
\end{eqnarray}
Since $\lambda_0$ must lie between $\pm i$,  
the $Q_f$-dependent part  of the denominator of 
equation (\ref{finaltdist}) can be approximated 
further and exponentiated to finally obtain a 
Gaussian distribution.  Hence it appears that in the large 
time limit, the central 
part of the distribution is  Gaussian and, as a consequence,  
the distribution satisfies the conventional
fluctuation theorem for small fluctuations. 

That the  asymptotic result  
in the large time limit indicates a Gaussian behavior 
 for small fluctuations is similar to what has been 
observed earlier through 
an explicit derivation 
 of the heat distribution function 
 for a uniformly moving 
 parabolic potential  \cite{zon,zon_cohen2}. 
 For large fluctuations,  	
 the distribution deviates from the  Gaussian one.
 The existence of a non-Gaussian tail in the 
heat  distribution for the same system  has been 
also  proved generally using the  energy 
conservation relation\cite{taniguchi1}.  
  Although,  we have not evaluated an  explicit 
 analytical form of the distribution,  under  the 
most general circumstances, 
 here also we expect similar non-Gaussian 
 feature for large fluctuations.  This inference is supported by 
 results from numerical integration and simulations discussed 
 below.

 \begin{figure}[htbp]
  \begin{center}
  (a) \includegraphics[width=.45\textwidth,clip,
angle=0]{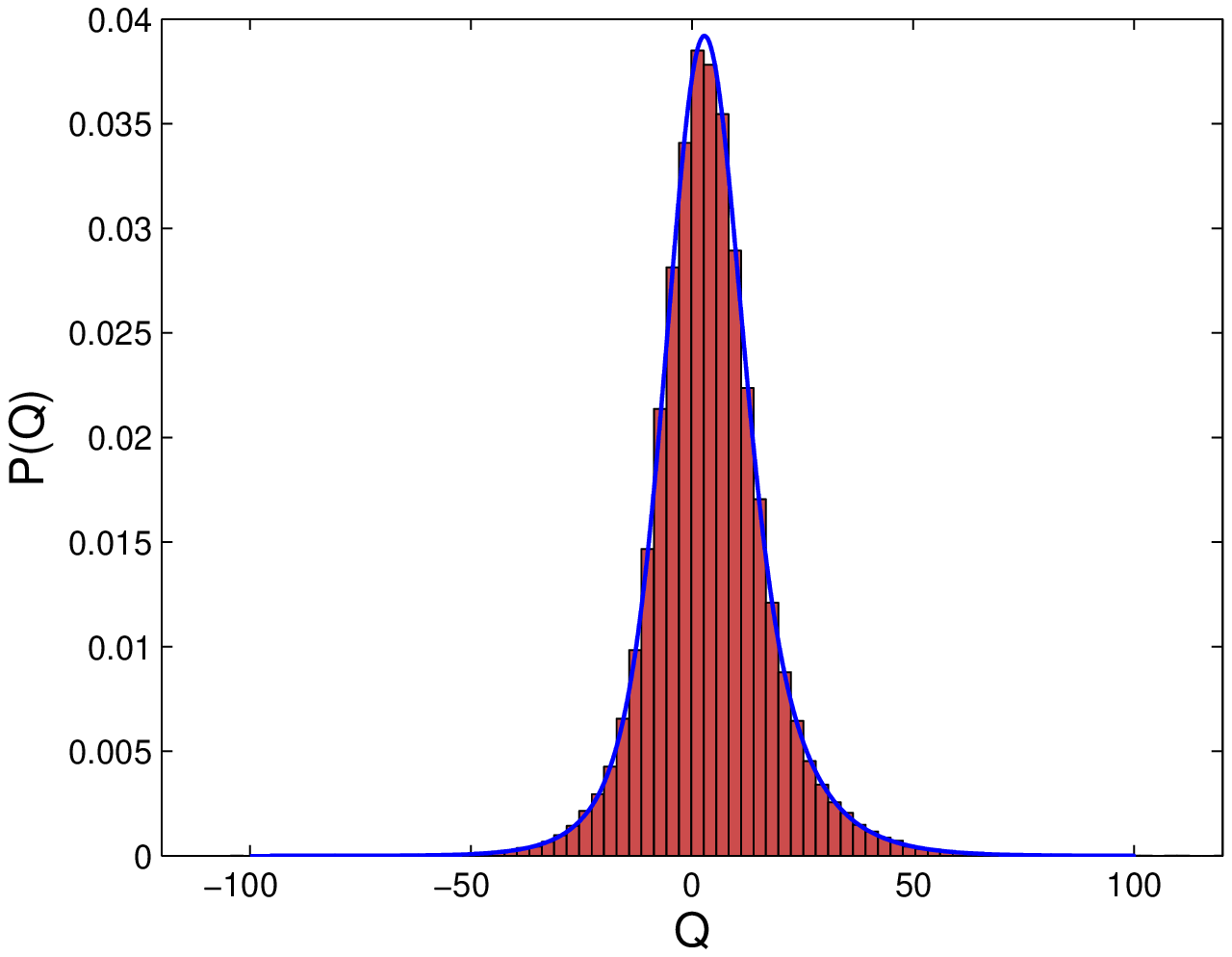}
(b)   \includegraphics[width=.45\textwidth,clip,
angle=0]{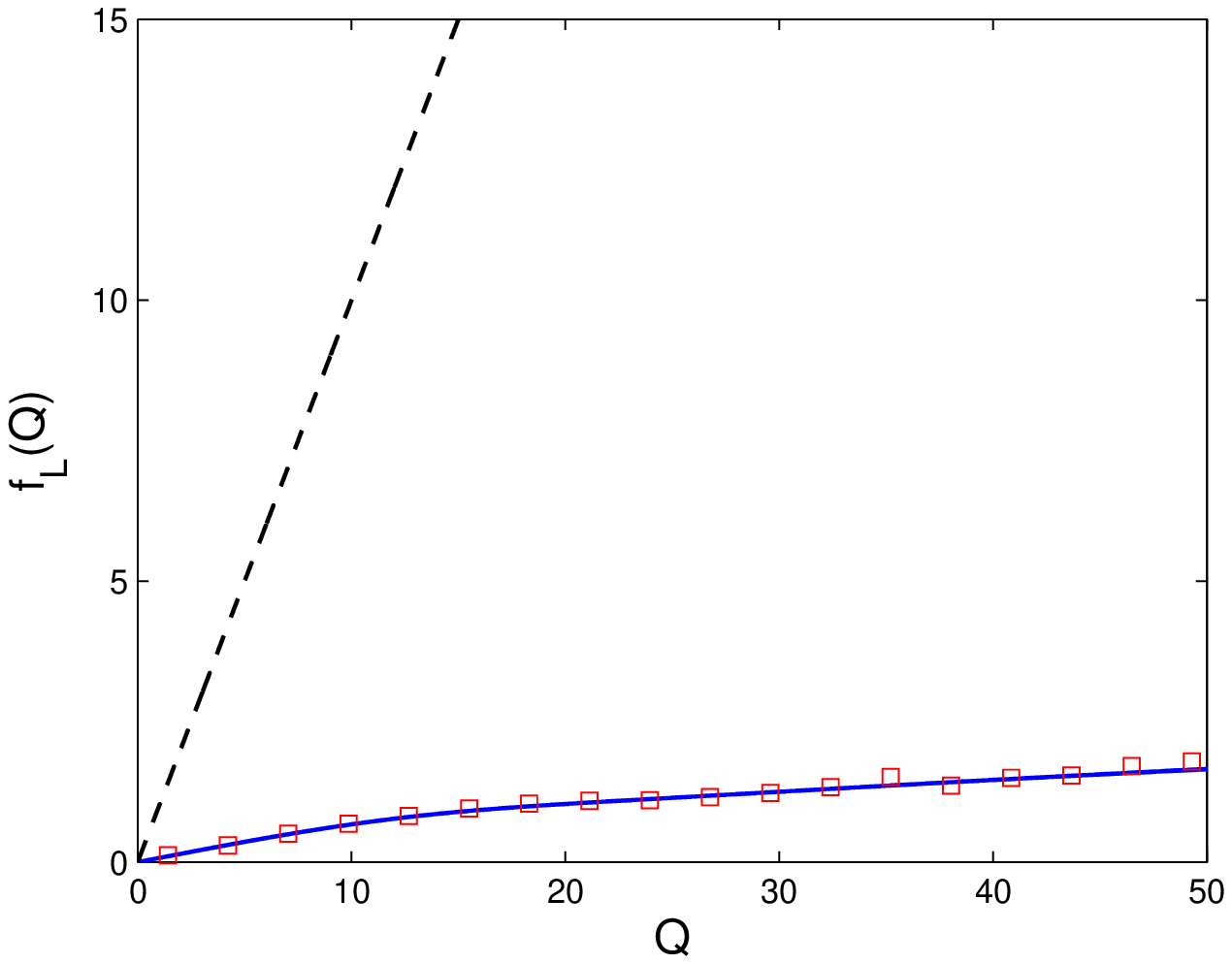}
\caption{(a) Distribution function of dissipated heat 
at time $t_f=10$. The histogram 
is obtained from numerical simulations after 
averaging over $10^5$ random 
trajectories. The solid blue line enveloping the histogram 
is obtained from the 
numerical integration of the    expression 
appearing on r.h.s. of 
equation(\ref{heatdist_ft}). Other parameter values are 
$D=0.05,\ \eta=0.05,\ \omega=0.1\pi,\ \gamma=0.1,\ \beta=0.1,\ \alpha=200$.  
(b) Plot of $\ln\left(\frac{P(Q)}{P(-Q)}\right)$ vs $Q$. 
The blue solid line is 
obtained from the numerical integration of the
 analytical expression (\ref{heatdist_ft}), 
and the red squares are obtained from numerical simulations. The black 
dashed line is the prediction of the fluctuation theorem. }
\label{fig:heatpdf}
  \end{center}
\end{figure}

Next, we evaluate the integration in equation
(\ref{heatdist_ft}) numerically using MATLAB which 
implements adaptive Gauss-Kronrod quadrature formula. 
 For the parameter values mentioned in figure(\ref{fig:heatpdf}-a), 
 we perform the numerical integration 
 for which the error bound is $1.48\times10^{-9}$. 
  The resulting distribution is plotted in 
 figure(\ref{fig:heatpdf}-a) which also shows a reasonably 
 good  agreement with the 
 histogram obtained from numerical simulations.
 Next, we consider the function 
$f_L(Q)=\ln\left(\frac{P(Q)}{P(-Q)}\right)$. 
To satisfy the transient fluctuation theorem, it is required that 
 $
 f_L(Q)=Q .
$  
In figure(\ref{fig:heatpdf}-b), we have plotted
 $f_L(Q)$ with $Q$ for $t_f=10$. From this plot, 
it is evident that  the probability distribution 
function  of the dissipated heat does not satisfy the transient 
fluctuation theorem.

\section{Numerical simulations}

Using the Euler-Maruyama method \cite{higham}, a discretized version  
of equation (\ref{langevin}) can be written as 
\begin{equation}
 x_{n+1}=x_n-\Delta t \ \eta \cos(\omega \ n\Delta t)-
\gamma x_n\Delta t+\sqrt{2D}\ 
{\rm d}\xi_n ,
\end{equation}
where $\Delta t=(t_f-t_i)/N$ is  a small time interval that divides the 
entire time interval $t_f-t_i$ into $N$  parts 
 and  $x_n$ denotes the position of the particle 
at $t_n=t_i+ \Delta t \ n$ with 
$n=0,1 \dots N$. ${\rm d}\xi_1$ \dots ${\rm d}\xi_n$ are independent 
normally distributed random variables with mean zero 
and standard deviation $\sqrt{\Delta t}$.
These random variables are generated using the 
in-built `random' function of MATLAB  that supplies
Gaussian-distributed 
random numbers ${\cal N}(0,1)$ with 
zero mean and unit standard deviation.

After calculating the position of the particle at each time-step, 
we have calculated the 
work and dissipated heat at each time step using 
equation (\ref{work}) and (\ref{heatdef}) 
respectively. To obtain the histogram for the  
work distribution, we have computed its values at final time  $t_f=1000$ 
 for $40000$ random trajectories with time step taken as 
$\Delta t = 0.05$. 
The histogram for the work probability
 distribution obtained 
from this data is shown in figure(\ref{fig:workdist_lt}-a). 
The variations 
of the standard deviation with the strength and the angular 
frequency 
of the oscillatory drive are  
  shown in figure (\ref{fig:workdist_lt}-b) and  (\ref{fig:sd_freq}-a), 
respectively. In figure (\ref{fig:sd_freq}-b), we have 
shown the variation 
  of the central fourth moment $\langle (W-\langle W\rangle)^4\rangle$ 
  and the variance of the work distribution with $\eta$ 
at large time $t_f=1000$.
  The agreement with the  relation \\
  $\langle (W-\langle W\rangle)^4\rangle = 
  3 (\langle W^2\rangle - \langle W\rangle^2)^2$  is also 
  shown in this figure. We have also seen the  variation 
of the coefficient of  skewness, 
  $\frac{\langle (W-\langle W\rangle)^3\rangle}{(\langle W^2\rangle - 
  \langle W\rangle^2)^{3/2}}$,  with  $\eta$. Over a range of  $\eta$ 
  between $[0.05:1]$,  the coefficient of skewness varies between $0.001$ 
  and $0.016$.

To plot the distribution 
of dissipated heat, we compute the heat values at 
time $t_f=10$  for $10^5$ realizations with the time 
step $\Delta t=0.0125$. The histogram is then compared 
with the analytical result in figure (\ref{fig:heatpdf}-a). 
In figure (\ref{fig:heatpdf}-b), we have compared the function 
$f_L(Q)$ obtained from numerical simulations with its 
values obtained after numerically evaluating the Fourier 
transform (\ref{heatdist_ft}).

\section{Summary} In this paper, using the Onsager-Machlup 
fluctuation theory, we  obtain the   
distribution functions  for work and dissipated 
heat for  a 
Brownian particle 
subjected to a confining harmonic potential and an oscillatory
 driving force.  We start with 
a Langevin equation  for the Brownian particle and obtain the transition 
probability in a  functional integral form with an Onsager-Machlup type 
Lagrangian  function.  The final 
form of the transition probability  is obtained by 
evaluating the  contribution from the 
optimal path and    deriving  
 the   functional integral explicitly for the  
fluctuations about the optimal path.  
 From  the Langevin equation and the energy conservation 
 principle, we identify  the total work  done on the Brownian particle. 
 This work consists of  two parts, 
one associated with the change in the potential 
 energy and the other, the heat dissipated to the reservoir.  
 The Onsager-Machlup type Lagrangian 
function can be expressed in terms of the 
entropy production rate and the dissipation functions. 
The  expressions for the  entropy production 
rate and the work done by the external force, obtained above,  are 
shown to be consistent with the energy conservation principle.

 The expression  for the  total work
is used  further to define  the work distribution function in a functional 
integral form.   As before,  
the  functional integral is  evaluated  by considering the 
optimal path and  the fluctuations about this. 
The work distribution function satisfies the  work fluctuation theorem 
in the long time limit.  While the 
width of the distribution saturates for large values of the angular 
 frequency of the oscillatory drive,   the width  
increases  with the angular frequency   for small values of the frequency.  
 Using the results derived in the 
work distribution part, we  obtain the 
Fourier transform of the distribution of the dissipated   heat.  
From this, using a saddle point approximation, 
 an  approximate analytical expression 
for the heat distribution is obtained in the long time limit. 
This result shows that for small fluctuations, the  resulting distribution 
is Gaussian. The 
heat distribution, at finite time, is obtained 
by numerically evaluating 
the inverse Fourier transform.  This result as well as the 
results from numerical 
simulations show that the heat distribution function does not 
satisfy transient fluctuation theorem.

\newpage
\appendix
\section{Calculation of the functional integral}\label{App:AppendixA}
In this appendix we evaluate the functional integral of equation 
(\ref{evaluatepathint}).
The time interval is split into $N$  slices through the following 
specifications $t_n=t_i+n \Delta t_N$ where $n=1,2,...,N$ 
and $\Delta t_N=(t_f-t_i)/N$.
The functional  integral can be written in the summation form as 
\begin{eqnarray}
&&\int {\cal D}z \ \exp\Big[-\frac{1}{4D}
\int_{t_i}^{t_f} {\rm d}t\ [\gamma z_t+\dot{z}_t]^2\Big]\nonumber\\
&& =\lim_{N\rightarrow \infty}
\Big(\frac{1}{4\pi D\Delta t_N}\Big)^{N/2} 
\int {\rm d}z_{N-1} {\rm d}z_{N-2} .... {\rm d}z_1 \exp\Big[-\frac{1}{4D}
\sum_{n=0}^{N-1}\Delta t_N \Big(\frac{(z_{n+1}-z_n)}{\Delta t_N}+
\gamma z_n\Big)^2\Big]\nonumber\\
&& =\lim_{N\rightarrow \infty}
\Big(\frac{1}{4\pi D\Delta t_N}\Big)^{N/2} 
\int {\rm d}z_{N-1} {\rm d}z_{N-2}.... {\rm d}z_1
\exp \Big[-\frac{1}{4D\Delta t_N}
\sum_{n=0}^{N-1} \Big(z_{n+1}+z_n \phi\Big)^2\Big],\label{evaluatepathint1}
\end{eqnarray}
where $\phi=\Delta t_N\gamma-1$. The Gaussian integral over 
all the $z$ variables can be done using the  matrix method. 
We however follow the straightforward method of completing square 
for  one integral at a time. Starting with  $z_{N-1}$ integral, 
we have 
\begin{eqnarray}
&&\lim_{N\rightarrow \infty}
\Big(\frac{1}{4\pi D\Delta t_N}\Big)^{N/2} 
\int {\rm d}z_{N-1} {\rm d}z_{N-2}.... {\rm d}z_1
\exp \Big[-\frac{1}{4D\Delta t_N}
\sum_{n=0}^{N-1} \Big(z_{n+1}+z_n \phi\Big)^2\Big]\nonumber\\
&& =\lim_{N\rightarrow \infty}
\Big(\frac{1}{4\pi D \Delta t_N}\Big)^{(N-1)/2} (1+\phi^2)^{-1/2}
\int  {\rm d}z_{N-2}.... {\rm d}z_1
\exp \Big[-\frac{1}{4D\Delta t_N}
\sum_{n=0}^{N-3} \Big(z_{n+1}+z_n \phi\Big)^2-\nonumber\\ 
&&\frac{1}{4D\Delta t_N} 
\frac{\phi^4}{1+\phi^2} z_{N-2}^2\Big] .
\end{eqnarray}
While doing this integration, we have used $z_N=0$. After 
performing integrations over $N-2$ variables starting with variable 
$z_{N-1}$, we are left with the final integration over $z_1$. 
(\ref{evaluatepathint1}) reduces to 
\begin{eqnarray}
&&\lim_{N\rightarrow \infty}
\Big(\frac{1}{4\pi D\Delta t_N}\Big)^{N/2} \int {\rm d}z_{N-1} 
{\rm d}z_{N-2}.... {\rm d}z_1
\exp \Big[-\frac{1}{4D\Delta t_N}
\sum_{n=0}^{N-1} \Big(z_{n+1}+z_n \phi\Big)^2\Big] \nonumber\\
&&=\lim_{N\rightarrow \infty} \Big( \frac{1}{4\pi D \Delta t_N}\Big)^{-1} 
(1+\phi^2+\phi^4+... 
+\phi^{2(N-2)})^{-1/2}\nonumber\\
&&\times \int {\rm d}z_1 \exp\Big[-\frac{1}{4D\Delta t_N} z_1^2-
\frac{1}{4D\Delta t_N} 
\frac{\phi^{2(N-1)}}{(1+\phi^2+\phi^4...+\phi^{2(N-2)})}z_1^2\Big].
\end{eqnarray}
The integration over $z_1$ leads to the final result 
\begin{eqnarray}
&&\lim_{N\rightarrow \infty}
\Big(\frac{1}{4\pi D\Delta t_N}\Big)^{N/2} 
\int {\rm d}z_{N-1} {\rm d}z_{N-2}.... {\rm d}z_1
\exp \Big[-\frac{1}{4D\Delta t_N}
\sum_{n=0}^{N-1} \Big(z_{n+1}+z_n \phi\Big)^2\Big]\nonumber\\
&&=\lim_{N\rightarrow \infty}\Big(\frac{1}{4\pi D \Delta t_N}\Big)^{1/2}
\Big[1+\phi^2+.... \phi^{2(N-1)}\Big]^{-1/2}\nonumber\\
&&=\lim_{N\rightarrow \infty}\Big(\frac{2\pi D}{\gamma}\Big)^{-1/2}
\Big(1-\frac{\gamma(t_f-t_i)}{2N}\Big)^{1/2}\Big[1-\Big(1-
\frac{\gamma(t_f-t_i)}{N}\Big)^{2N}\Big]^{-1/2}.
\end{eqnarray}
In order to obtain the last step, we have used 
\begin{eqnarray}
\left(1+\phi^2+\phi^4+...\phi^{2(N-1)}\right)^{-1/2}=
\Big[\frac{(1-\phi^2)}{(1-\phi^{2N})}\Big]^{1/2}.
\end{eqnarray}
Next, we need to consider  $N\rightarrow \infty$ limit. 
This finally leads to 
\begin{eqnarray}
\int {\cal D}z \ \exp\Big[-\frac{1}{4D}\int_{t_i}^{t_f} {\rm d}t\ 
(\gamma z_t+
\dot{z}_t)^2\Big]
=\Big(\frac{2\pi D}{\gamma}\Big)^{-1/2}
\Big(1-\exp[-2\gamma(t_f-t_i)]\Big)^{-1/2}.	 
\end{eqnarray}

\section{Expressions appearing in the various 
integrations in section  \ref{sec:workd} and 
\ref{sec:heatd}}\label{App:AppendixB}
The expressions for $c_1$, $c_2$ and $c_3$ in 
equation(\ref{beforexfint}) are 
\begin{eqnarray}
&&c_1 =\frac{2\gamma e^{2\gamma t_f}}{(e^{2\gamma t_f}-1)}.\\
&&c_2 =-4\gamma \left[ \frac{x_f \exp(\gamma t_f)}
{(\exp(2\gamma t_f)-1)} -A_W\right] 
e^{\gamma t_f}+
\frac{2i\bar{\lambda}}{(\omega^2+\gamma^2)}\bar{\omega}
\end{eqnarray}
and  
\begin{eqnarray}
&&c_3=-2\gamma \left[ \frac{x_f \exp(\gamma t_f)}
 {\exp(2\gamma t_f)-1}-A_W\right]^2(1-\exp(2\gamma t_f))-
\frac{\gamma\bar{\lambda^2}}{2(\omega^2+\gamma^2)^2}+
\frac{2i\bar{\lambda}\eta\gamma}{4\omega (\omega^2+\gamma^2)}-\nonumber\\
&&\frac{2i\bar{\lambda}\omega}{(\omega^2+\gamma^2)}
\left[x_i+\frac{\eta\gamma}{(\omega^2+\gamma^2)}\right]
+\frac{\bar{\lambda}^2 t_f}{2(\omega^2+\gamma^2)}+
\frac{i\bar{\lambda}\eta\omega t_f}{(\omega^2+\gamma^2)}-
\frac{2i\bar{\lambda} \gamma }{(\omega^2+\gamma^2)} 
\left[ \frac{x_f \exp(\gamma t_f)} {\exp(2\gamma t_f)-1}-A_W\right]
\exp(\gamma t_f)\sin\omega t_f-\nonumber\\
&&\left [\frac{{\bar{\lambda}}^2
(\omega^2-\gamma^2)}{4\omega (\omega^2+\gamma^2)^2}
+\frac{2i\bar{\lambda}(\eta\omega-i\bar{\lambda})}
{4\omega(\omega^2+\gamma^2)} \right]\sin 2\omega t_f+
\frac{2i\bar{\lambda}\bar{\omega}\exp(-\gamma t_f)}{(\omega^2+\gamma^2)}
\left[ x_i+
\frac{\eta\gamma}{(\omega^2+\gamma^2)}+
\frac{x_f \exp(\gamma t_f)}{(\exp(2\gamma t_f)-1)}-A_W\right]\nonumber\\
&&-\frac{2i\bar{\lambda}\omega}{(\omega^2+\gamma^2)} \exp(\gamma t_f)
 \left[\frac{x_f\exp(\gamma t_f)}{(\exp(2\gamma t_f)-1)}-A_W\right]
\cos \omega t_f+
\frac{1}{4\omega} \left[\frac{2\gamma \omega \bar{\lambda}^2}
{(\omega^2+\gamma^2)^2}-
\frac{2i\bar{\lambda}\eta\gamma}{(\omega^2+\gamma^2)}\right]\cos 2\omega t_f,
\end{eqnarray}
where $\bar{\omega}=\omega\cos\omega t_f+\gamma \sin\omega t_f$ .

The expressions for $m_1$ and $m_2$ in equation 
(\ref{m1m2}) of the main text are
\begin{eqnarray}
&& m_2=-[\gamma^2 \omega (1+e^{2\gamma t_f})+\omega^3(1-e^{2\gamma t_f})
-2e^{2\gamma t_f} \gamma \omega t_f(\gamma^2+\omega^2)-
\gamma^2 \omega \cos(2\omega t_f)(1+e^{2\gamma t_f})\nonumber\\
&& +\omega^3 \cos(2\omega t_f)(1-e^{2\gamma t_f})+
2\gamma \omega^2 \sin(2\omega t_f)+
e^{2\gamma t_f} \gamma (\gamma^2+\omega^2)\sin(2\omega t_f)]
(16D e^{2\gamma t_f}\gamma\omega (\gamma^2+\omega^2)^2)^{-1}
\end{eqnarray}
and 
\begin{eqnarray}
m_1=-\frac{i}{8}\left[\eta\gamma(\cos 2\omega t_f-1)+(x_i+
\frac{\eta\gamma}{(\omega^2+\gamma^2)})4\omega
(\omega- \bar{\omega} \exp[-\gamma t_f])-
2\omega^2\eta t_f+\eta\omega \sin 2\omega t_f\right]
(D \omega(\gamma^2+\omega^2))^{-1}
\end{eqnarray}

The expressions for  $l_4,l_3,l_2,l_1,l_0$ 
appearing in equation(\ref{exprqn}) are,
\begin{eqnarray}
l_4&=& -2\eta^2\gamma\omega \biggl\{ (1+e^{2\gamma t_f})
\gamma\omega\cos(2\omega t_f)+e^{\gamma t_f}\Big[ -
2\gamma\omega \cosh(\gamma t_f)+2\omega t_f\sinh(\gamma t_f)
(\omega^2+\gamma^2)\nonumber \\
&&+(\omega^2-\gamma^2)\sin(2\omega t_f)\sinh(\gamma t_f)\Big] \biggr\}.
\end{eqnarray}

\begin{eqnarray}
l_3&=& -2\eta^2\gamma\omega \biggl\{ (1+e^{2\gamma t_f})
\gamma\omega\cos(2\omega t_f)+e^{\gamma t_f}\Big[ -
2\gamma\omega \cosh(\gamma t_f)+2\omega t_f\sinh(\gamma t_f)
(\omega^2+\gamma^2)\nonumber \\
&&+(\omega^2-\gamma^2)\sin(2\omega t_f)\sinh(\gamma t_f)\Big] \biggr\}.
\end{eqnarray}

\begin{eqnarray}
l_2&=&\eta^2\gamma \biggl\{ -2\gamma(\gamma^2+2\omega^2)
 +8e^{\gamma t_f}\omega^3\sin(\omega t_f)+e^{2\gamma t_f}
\Big[ 2\gamma(\gamma^2+3\omega^2)-2\omega^2
(\omega^2+\gamma^2)t_f\nonumber\\
&&-2\gamma\omega^2\cos(2\omega t_f)+(\gamma^2-\omega^2)
\omega\sin(2\omega t_f)\Big] \biggr\}.
\end{eqnarray}

\begin{eqnarray}
l_1&=& \eta^2\omega \biggl\{ 2\omega^3 +
8e^{\gamma t_f}\gamma\omega^2\sin(\omega t_f)-e^{2\gamma t_f}
\Big[ -2\omega\gamma^2+2\omega^3+ 2\omega\gamma t_f
(\omega^2+\gamma^2)\nonumber \\
&&+2\omega\gamma^2\cos(2\omega t_f)+\gamma(\omega^2-\gamma^2)
\sin(2\omega t_f)\Big] \biggr\}.
\end{eqnarray}

and
\begin{eqnarray}
l_0=2e^{2\gamma t_f}\eta^2(\omega^2+\gamma^2)^2.
\end{eqnarray}


\begin{thebibliography}{99}
\bibitem{most-probable} For a linear model, an approximation through most probable 
  path is expected to be  a good approximation in the small $D$ limit \cite{wiegel}.
\end{thebibliography}
\end{document}